\documentclass[10pt]{article}
\usepackage{url, graphicx, color, varioref, amsfonts, longtable, float,epsfig,amsthm}
\title{Hiding Quantum States in a Superposition}

\author{Ahmed Younes\footnote {ayounes2@yahoo.com}\\
Faculty of Science\\
Alexandria University\\
Alexandria, Egypt}

\begin{document}
\maketitle
\begin{abstract}
A method to hide certain quantum states in a superposition will be proposed. 
Such method can be used to increase the security of a communication channel. 
States represent an encrypted message will disappear during data exchange. 
This makes the message 100$\%$ safe under direct measurement by an eavesdropper.
No entanglement sharing is required among the communicating parties.

\end{abstract}

%\tableofcontents

\section{Introduction}

Establishing a secure channel is of interest. There have been efforts to provide a way 
to exploit quantum mechanical systems to achieve this goal. For example, quantum key distribution 
\cite{Bennett92,Ekert92,Lo99} has been there for some time. Methods to 
hide classical data using quantum states among parties \cite{Div01,Terhal00}. The aim of this paper 
is to show how to remove quantum states that carry encrypted message from a superposition. The proposed 
method is not a substitution of the above efforts. It can be used together with any encryption algorithm 
to increase the securing of the channel during communication. 

The paper is organized as follows: 
Section \ref{sec2} introduces the basic idea of hiding quantum states.
Section \ref{sec3} explains the process of hiding certain quantum states.
Section \ref{sec4} shows how to unhide hidden states in the superposition.
Section \ref{sec5} gives the analysis of the communication in the presence of eavesdropping.
Section \ref{sec6} proposes the transmission protocol by hiding a message.
Section \ref{sec7} suggests possible variations to the transmission protocol to enhance the security level.
The paper ends up with a conclusion in Section \ref{sec8}.

\section{Basic Idea}
\label{sec2}
Suppose that Alice and Bob want to exchange a message in a secure way. Alice and Bob 
can use any encryption algorithm to encode the stream of bits. They 
exchange the message one bit at a time using two quantum wires. Assume the following 2-qubit quantum system,

\begin{equation}
\left| \psi  \right\rangle  = \frac{1}{2}\left( {\left| {00} \right\rangle  + \left| {01} \right\rangle  + \left| {10} \right\rangle  + \left| {11} \right\rangle } \right).
\label{eq1}
\end{equation}

The aim is to hide any two quantum states from the system, so $\left| \psi  \right\rangle$ in Eqn.~(\ref{eq1}) to be converted to 
one of the following systems,

\begin{equation}
\begin{array}{*{20}c}
   {\frac{1}{{\sqrt 2 }} \left( {\left| 00 \right\rangle  + \left| 01 \right\rangle } \right),} & {\frac{1}{{\sqrt 2 }}\left( {\left| 10 \right\rangle  + \left| 11 \right\rangle } \right),} & {\frac{1}{{\sqrt 2 }}\left( {\left| {00} \right\rangle  + \left| {11} \right\rangle } \right),}  \\
   {\frac{1}{{\sqrt 2 }} \left( {\left| 00 \right\rangle  + \left| 10 \right\rangle } \right),} & {\frac{1}{{\sqrt 2 }}\left( {\left| 01 \right\rangle  + \left| 11 \right\rangle } \right),} & {\frac{1}{{\sqrt 2 }}\left( {\left| {01} \right\rangle  + \left| {10} \right\rangle } \right).}  \\
\end{array}
\end{equation}

This is useful in data exchange. To see this, assume that Alice wants to send 0 to Bob, 
Alice has two choices: to use the first qubit so the superposition takes the form 
${\textstyle{1 \over {\sqrt 2 }}}\left( {\left| {00} \right\rangle  + \left| {01} \right\rangle } \right)$, 
or to use the second qubit so the superposition takes the form 
${\textstyle{1 \over {\sqrt 2 }}}\left( {\left| {00} \right\rangle  + \left| {10} \right\rangle } \right)$, 
i.e. one qubit is used to encode the data and the other qubit is used to confuse any eavesdropper. 
In this case, the message is 50$\%$ secure under any direct measurement by the eavesdropper.
To increase the security of the message, Alice then has another two choices, 
to hide or not to hide the states. Hiding the states is to convert the above prepared system to
${\textstyle{1 \over {\sqrt 2 }}}\left( {\left| {11} \right\rangle  + \left| {10} \right\rangle } \right)$ or
${\textstyle{1 \over {\sqrt 2 }}}\left( {\left| {11} \right\rangle  + \left| {01} \right\rangle } \right)$ 
respectively. In this case, the message is 100$\%$ secure under any direct measurement. 
Direct measurement will lead to states that don't represent the required message.

\section{Hiding Quantum States}
\label{sec3}
The aim of this section is to show the process of selecting then (optionally) hiding certain quantum states 
from the superposition shown in Eqn.~(\ref{eq1}). To achieve this, assume that we have two quantum operators: 
$U_f$ and $D_p$, and a 3-qubit system, where the third qubit is initialized to $\left|0\right\rangle$ 
and will be used temporarily to mark the required states via entanglement. $U_f$ is an oracle that evaluates 
to 1 only for the required two states. $D_p$ is an operator which performs the inversion about the mean, 
and a phase shift of -1 on the {\it subspace} 
of the system entangled with the extra qubit in state $\left|0\right\rangle$, 
and $\left|1\right\rangle$ respectively.

The diagonal representation of $D_p$ when applied on $3$-qubit system can take this form \cite{Younes07},

\begin{equation}
\label{ENheq13}
D_p =  \left(W^{ \otimes 2}  \otimes I_1\right)\left( {2\left| 0 \right\rangle \left\langle 0 \right| - I_{3}} \right)\left(W^{ \otimes 2}  \otimes I_1\right),
\end{equation}
\noindent
where the vector $\left| 0 \right\rangle$ used in Eqn. \ref{ENheq13} is of length 8, $I_k$ is the identity matrix 
of size $2^k\times 2^k$, and $W$ is the Hadamard gate. Consider a general state $\left|\psi\right\rangle$ of $3$-qubit register:

\begin{equation}
\begin{array}{l}
\left| \psi  \right\rangle  = \sum\limits_{k = 0}^{7} {\delta _k \left| k \right\rangle } = \sum\limits_{j = 0}^{3} {\alpha _j \left( {\left| j \right\rangle  \otimes \left| 0 \right\rangle } \right)}  + \sum\limits_{j = 0}^{3} {\beta _j \left( {\left| j \right\rangle  \otimes \left| 1 \right\rangle } \right)},
\end{array}
\end{equation}

%For our purposes and without loss of generality, the general system $\left|\psi\right\rangle$ can be re-written as,

%\begin{equation}
%\label{ENheq14}
%\left| \psi  \right\rangle 
%\end{equation}

\noindent
where \{$\alpha _j  = \delta _k$ : $k$ even\} and \{$\beta _j  = \delta _k$ : $k$ odd\}. 
The effect of applying $D_p$ on $\left| \psi  \right\rangle$ produces,

\begin{equation}
\label{ENheq15}
\begin{array}{l}
D_p\left| \psi  \right\rangle = \left( W^{ \otimes 2}  \otimes I_1 \right) \left( {2\left| 0 \right\rangle \left\langle 0 \right| - I_{3}} \right) \left( W^{ \otimes 2}  \otimes I_1 \right)\sum\limits_{k = 0}^{7} {\delta _k \left| k \right\rangle }\\
\,\,\,\,\,\,\,\,\,\,\,\,\,\,\,\,\,\, = 2\left( {W^{ \otimes 2}  \otimes I_1\left| 0 \right\rangle \left\langle 0 \right|W^{ \otimes 2}  \otimes I_1} \right)\sum\limits_{k = 0}^{7} {\delta _k \left| k \right\rangle }  - \sum\limits_{k = 0}^{7} {\delta _k \left| k \right\rangle }\\
\,\,\,\,\,\,\,\,\,\,\,\,\,\,\,\,\,\, =\sum\limits_{j = 0}^{3} {2\left\langle \alpha  \right\rangle \left( {\left| j \right\rangle  \otimes \left| 0 \right\rangle } \right)}  - \sum\limits_{k = 0}^{7} {\delta _k } \left| k \right\rangle\\ 
\,\,\,\,\,\,\,\,\,\,\,\,\,\,\,\,\,\,=\sum\limits_{j = 0}^{3} {\left( {2\left\langle \alpha  \right\rangle  - \alpha _j } \right)\left( {\left| j \right\rangle  \otimes \left| 0 \right\rangle } \right)}  - \sum\limits_{j = 0}^{3} {\beta _j \left( {\left| j \right\rangle  \otimes \left| 1 \right\rangle } \right)},\\
\end{array}
\end{equation}

\noindent
where $\left\langle \alpha  \right\rangle  = \frac{1}{4}\sum\nolimits_{j = 0}^{3} {\alpha _j }$ is the mean of 
the amplitudes of the subspace ${\alpha _j \left( {\left| j \right\rangle  \otimes 
\left| 0 \right\rangle } \right)}$, i.e. applying the operator $D_p$ will perform the inversion about the mean 
only on the subspace ${\alpha _j \left( {\left| j \right\rangle  \otimes 
\left| 0 \right\rangle } \right)}$ and will only {\it change the sign} of the amplitudes for the subspace 
${\beta _j \left( {\left| j \right\rangle  \otimes \left| 1 \right\rangle } \right)}$.  

The purpose of using $D_p$ is to hide certain states in the superposition. To see this, 
let $U_f$ be an oracle that has the following property:
\begin{equation}
U_f \left| {x,0} \right\rangle  \to \left| {x,f(x)} \right\rangle,
\end{equation}

and
\begin{equation}
\left| \psi  \right\rangle  = \frac{1}{2}\left( {\left| {x_0 } \right\rangle  + \left| {x_1 } \right\rangle  + \left| {x_2 } \right\rangle  + \left| {x_3 } \right\rangle } \right) \otimes \left| 0 \right\rangle.
\end{equation}

Assume that $U_f$ evaluates to 1 for any arbitrary two states in the superposition, i.e.

\begin{equation}
U_f \left| \psi  \right\rangle  = \frac{1}{2}\left( {\left| {x_0 } \right\rangle  + \left| {x_1 } \right\rangle } \right) \otimes \left| 0 \right\rangle  + \frac{1}{2}\left( {\left| {x_2 } \right\rangle  + \left| {x_3 } \right\rangle } \right) \otimes \left| 1 \right\rangle,
\end{equation}
\noindent
where $x_i\in\{00,01,10,11\}$, 
then $\left\langle \alpha  \right\rangle  = {\textstyle{1 \over 4}}$ and 
$D_p U_f \left| \psi  \right\rangle$ will make the superposition contains 
only the states that make $U_f$ evaluates to 1, i.e. 
$\frac{1}{{\sqrt 2 }}\left( {\left| {x_2 } \right\rangle  + \left| {x_3 } \right\rangle } \right)$. 
To hide the states ${\left| {x_2 } \right\rangle }$ and ${\left| {x_3 } \right\rangle }$, 
then $U_{\overline f}$ will be used instead of $U_f$, where ${\overline f}=f\oplus1$. 
The extra qubit can then be omitted from the system.

\section{Showing Hidden States}
\label{sec4}
To show the hidden states with no prior knowledge of the oracle used, 
the operator $G$ used in the original Grover's algorithm \cite{grover96} 
to perform the usual inversion about the mean will be used once. 
The diagonal representation of $G$ on 2-qubit system can take this form,

\begin{equation}
\label{SInteqn6}
G = W^{ \otimes 2 }\left( {2\left| 0 \right\rangle \left\langle 0 \right| 
- I_2} \right)W^{ \otimes 2},
\end{equation}

\noindent
where the vector $\left|0\right\rangle$ used in Eqn. \ref{SInteqn6} is of length 4, and 
$I_{2}$ is the identity matrix of size $4\times 4$. 
Consider a general system $\left|\psi\right\rangle$ of $2$-qubit register:

\begin{equation}
\left| \psi  \right\rangle  = \sum\limits_{j = 0}^{3} {\alpha _j \left| j \right\rangle }.
\end{equation}

The effect of applying $G$ on $\left| \psi  \right\rangle$ gives,

\begin{equation}
G\left| \psi  \right\rangle  = \sum\limits_{j = 0}^{3} {\left[ { - \alpha _j  + 2\left\langle \alpha  \right\rangle } \right]\left| j \right\rangle },
\end{equation}
where, $\left\langle \alpha \right\rangle = \frac{1}{4}\sum\nolimits_{j = 0}^{3} {\alpha _j }$ 
is the mean of the amplitudes of the states in the superposition, 
i.e. each amplitude $\alpha _j $ will be transformed according to the following relation:

\begin{equation}
\label{SInteqn7}
\alpha _j \to \left[ { - \alpha _j + 2\left\langle \alpha \right\rangle } 
\right].
\end{equation}

To understand the purpose of using $G$, consider the following cases:
\begin{itemize}
\item[1-] If the system is in the form
\begin{equation}
\left| \psi  \right\rangle  = \frac{1}{2}\left( {\left| {00} \right\rangle  + \left| {01} \right\rangle  + \left| {10} \right\rangle  + \left| {11} \right\rangle } \right),
\end{equation}
\noindent
then $\left\langle \alpha  \right\rangle  = {\textstyle{1 \over 2}}$ and applying 
$G$ has no effect on the system.

\item[2-] If the system is in the form,
\begin{equation}
\left| \psi  \right\rangle  = \left| x \right\rangle, 
\end{equation}
\noindent
such that $x\in\{00,01,10,11\}$ then 
$\left\langle \alpha  \right\rangle  = {\textstyle{1 \over 4}}$ and applying $G$ on the system 
will create a superposition of all possible states with a phase shift of -1 on $\left| x \right\rangle$. 

\item[3-] If the system is in the form,
\begin{equation}
\left| \psi  \right\rangle  = \frac{1}{{\sqrt 2 }}\left( {\left| x \right\rangle  + \left| y \right\rangle } \right),
\end{equation}
\noindent
such that $x,y\in\{00,01,10,11\}$ then 
$\left\langle \alpha  \right\rangle  = {\textstyle{1 \over {2\sqrt 2 }}}$ and applying $G$ on the system 
will transfer the amplitudes to the states that don't exist in the superposition, i.e. any hidden 
states in the superposition will appear and already existing states will disappear.
\end{itemize}

\section{Transmission Protocol}
\label{sec5}

To send a bit, Alice prepares the system 
$D_p U_f \left| {00} \right\rangle  \otimes \left| 0 \right\rangle$. 
Alice chooses the appropriate $U_f$ according to the following:
\begin{itemize}
\item[1-] The data to be sent is 0 or 1.
\item[2-] The data will be placed on the first or the second qubit.
\item[3-] To hide or not to hide the states.

\end{itemize}

Alice chooses an oracle from eight possible oracles. 
The oracle represents both the message and the action applied by Alice on the message. 
Bob should know in advance
the correct qubit and that Alice will hide or not the states, but has no knowledge of the content of 
the message. Assume that Alice and Bob pre-agreed on a secret key of the form $(A,p)$, for example,

\begin{equation}
H1H2N1H2N2...,
\end{equation}
\noindent
where $A$ is an action such that $A \in \{H,N\}$, 
$H$ is to hide the superposition, $N$ is not to hide the superposition,
and $p$ is the bit position such that $p\in\{1,2\}$. According to the key, if the action is $H$ then Bob applies $G$ then reads the data from the 
appropriate qubit. If the action is $N$ then Bob directly reads data from the qubit. In case of 
no eavesdropping, Bob will get the message correctly.

\section{Detection of Eavesdropping}
\label{sec6}
Assume that Eve is trying to intercept the message. Eve neither knows the action applied by Alice nor the appropriate 
position of the message. Eve decides randomly to apply G then measure (action denoted by $GM$) or to directly apply measurement (action denoted by $M$). 
Then Eve has to decide randomly if the data is placed on the first or the second qubit. 
Eve has a chance of 25$\%$ to get the correct message. If the message is already encrypted, then Eve has 
another problem to decrypt the detected message correctly. Eve has to resend a superposition to Bob. Eve has two choices:
\begin{itemize}
\item[1-] To resend the measured data as it is, action denoted by $SM$.
\item[2-] To prepare a random superposition, action denoted by $PS$.
\end{itemize}  

To calculate the chance that Bob gets the correct message in the presence of eavesdropping, consider 
the following cases:
\begin{itemize}
	\item[Case 1:] 
	\begin{itemize}
	\item[1-] In case Eve applied $M$ and Bob was supposed to apply 
	$M$ then Bob will get the correct message with a chance of 100$\%$.

	\item[2-]  In case Eve applied $M$ and Bob was supposed to apply $GM$. 
	Applying $G$ by Bob will create all the possible 
	2-qubit states equally weighted, so Bob will get a random state with probability 25$\%$ and the 
	probability to get the data in the correct position is 50$\%$. Bob has a change of 12.5$\%$ to 
	get the correct message.
	\end{itemize}

	\item[Case 2:]
	In case Eve decides to prepare a superposition then she should decide: 
	(1) if the correct data is 0 or 1.
	(2) to place the data on the first or second qubit. 
	(3) to hide or not the states. 
	(4) to choose one oracle out of four possible oracles accordingly.
	Eve has a chance of 12.5$\%$ to prepare the correct superposition.
	
\end{itemize}  
 
\begin{center}
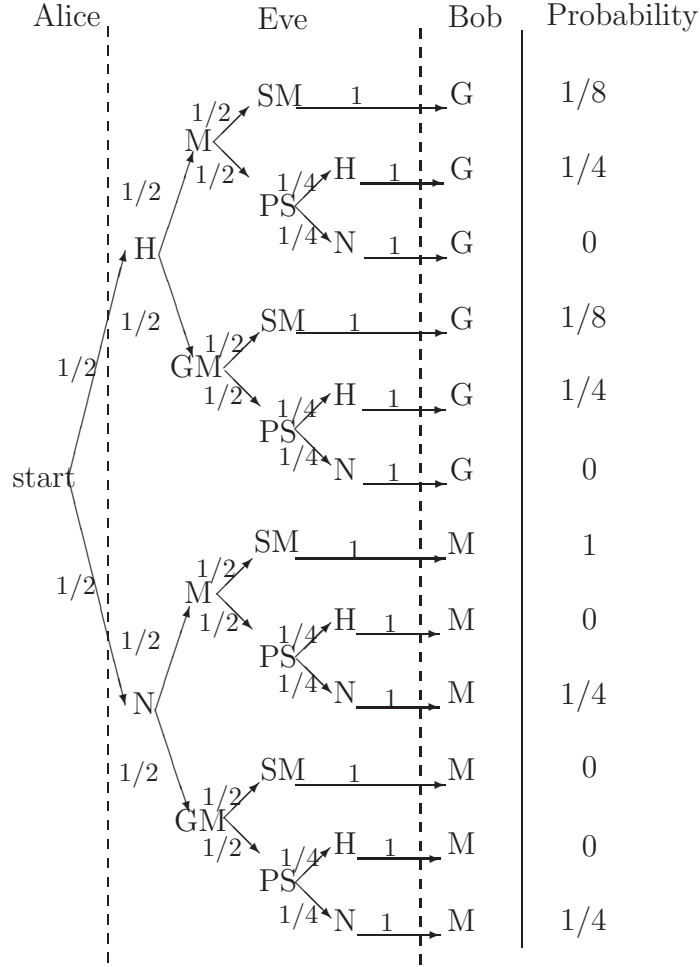
\begin{figure}  [htbp]
\begin{center}

\setlength{\unitlength}{3947sp}%
\begingroup\makeatletter\ifx\SetFigFont\undefined%
\gdef\SetFigFont#1#2#3#4#5{%
  \reset@font\fontsize{#1}{#2pt}%
  \fontfamily{#3}\fontseries{#4}\fontshape{#5}%
  \selectfont}%
\fi\endgroup%
\begin{picture}(3607,6139)(96,-5528)
\thinlines
{\put(1896,-135){\vector( 1, 0){939}}
}%
{\put(1896,-1551){\vector( 1, 0){939}}
}%
{\put(1890,-2966){\vector( 1, 0){939}}
}%
{\put(1890,-4388){\vector( 1, 0){939}}
}%
{\put(2299,-609){\vector( 1, 0){519}}
}%
{\put(2322,-1077){\vector( 1, 0){519}}
}%
{\put(2310,-2031){\vector( 1, 0){519}}
}%
{\put(2316,-2498){\vector( 1, 0){519}}
}%
{\put(2275,-3442){\vector( 1, 0){519}}
}%
{\put(2275,-3903){\vector( 1, 0){519}}
}%
{\put(2281,-4852){\vector( 1, 0){519}}
}%
{\put(2281,-5332){\vector( 1, 0){519}}
}%
{\put(1376,-348){\vector( 1, 1){222}}
}%
{\put(1439,-1769){\vector( 1, 1){222}}
}%
{\put(1396,-3187){\vector( 1, 1){222}}
}%
{\put(1439,-4595){\vector( 1, 1){222}}
}%
{\put(1447,-1786){\vector( 1,-1){222}}
}%
{\put(1375,-348){\vector( 1,-1){222}}
}%
{\put(1395,-3183){\vector( 1,-1){222}}
}%
{\put(1442,-4593){\vector( 1,-1){222}}
}%
{\put(1888,-750){\vector( 1, 1){222}}
}%
{\put(1889,-751){\vector( 1,-1){222}}
}%
{\put(1885,-2153){\vector( 1, 1){222}}
}%
{\put(1889,-2153){\vector( 1,-1){222}}
}%
{\put(1890,-3588){\vector( 1, 1){222}}
}%
{\put(1895,-3592){\vector( 1,-1){222}}
}%
{\put(1886,-5006){\vector( 1, 1){222}}
}%
{\put(1891,-5006){\vector( 1,-1){222}}
}%
{\put(3313,355){\line( 0,-1){5767}}
}%
{\put(461,-2467){\vector( 1, 4){360}}
}%
{\put(461,-2448){\vector( 1,-4){360}}
}%
{\multiput(2676,364)(0.00000,-118.78788){50}{\line( 0,-1){ 59.394}}
}%
{\multiput(713,377)(0.00000,-118.78788){50}{\line( 0,-1){ 59.394}}
}%
{\put(1034,-1053){\vector( 1, 3){216}}
}%
{\put(1010,-3915){\vector( 1, 3){216}}
}%
{\put(1032,-1055){\vector( 1,-3){216}}
}%
{\put(1008,-3914){\vector( 1,-3){216}}
}%
\put(2860,-106){\makebox(0,0)[lb]{\smash{{\SetFigFont{12}{14.4}{\rmdefault}{\mddefault}{\updefault}{G}%
}}}}
\put(2860,-580){\makebox(0,0)[lb]{\smash{{\SetFigFont{12}{14.4}{\rmdefault}{\mddefault}{\updefault}{G}%
}}}}
\put(2860,-1048){\makebox(0,0)[lb]{\smash{{\SetFigFont{12}{14.4}{\rmdefault}{\mddefault}{\updefault}{G}%
}}}}
\put(2860,-1522){\makebox(0,0)[lb]{\smash{{\SetFigFont{12}{14.4}{\rmdefault}{\mddefault}{\updefault}{G}%
}}}}
\put(2860,-2002){\makebox(0,0)[lb]{\smash{{\SetFigFont{12}{14.4}{\rmdefault}{\mddefault}{\updefault}{G}%
}}}}
\put(2860,-2469){\makebox(0,0)[lb]{\smash{{\SetFigFont{12}{14.4}{\rmdefault}{\mddefault}{\updefault}{G}%
}}}}
\put(2841,-2937){\makebox(0,0)[lb]{\smash{{\SetFigFont{12}{14.4}{\rmdefault}{\mddefault}{\updefault}{M}%
}}}}
\put(2841,-3413){\makebox(0,0)[lb]{\smash{{\SetFigFont{12}{14.4}{\rmdefault}{\mddefault}{\updefault}{M}%
}}}}
\put(2841,-3874){\makebox(0,0)[lb]{\smash{{\SetFigFont{12}{14.4}{\rmdefault}{\mddefault}{\updefault}{M}%
}}}}
\put(2841,-4359){\makebox(0,0)[lb]{\smash{{\SetFigFont{12}{14.4}{\rmdefault}{\mddefault}{\updefault}{M}%
}}}}
\put(2841,-4823){\makebox(0,0)[lb]{\smash{{\SetFigFont{12}{14.4}{\rmdefault}{\mddefault}{\updefault}{M}%
}}}}
\put(2841,-5303){\makebox(0,0)[lb]{\smash{{\SetFigFont{12}{14.4}{\rmdefault}{\mddefault}{\updefault}{M}%
}}}}
\put(2127,-580){\makebox(0,0)[lb]{\smash{{\SetFigFont{12}{14.4}{\rmdefault}{\mddefault}{\updefault}{H}%
}}}}
\put(2127,-1048){\makebox(0,0)[lb]{\smash{{\SetFigFont{12}{14.4}{\rmdefault}{\mddefault}{\updefault}{N}%
}}}}
\put(2127,-2002){\makebox(0,0)[lb]{\smash{{\SetFigFont{12}{14.4}{\rmdefault}{\mddefault}{\updefault}{H}%
}}}}
\put(2127,-2469){\makebox(0,0)[lb]{\smash{{\SetFigFont{12}{14.4}{\rmdefault}{\mddefault}{\updefault}{N}%
}}}}
\put(2127,-3413){\makebox(0,0)[lb]{\smash{{\SetFigFont{12}{14.4}{\rmdefault}{\mddefault}{\updefault}{H}%
}}}}
\put(2127,-3874){\makebox(0,0)[lb]{\smash{{\SetFigFont{12}{14.4}{\rmdefault}{\mddefault}{\updefault}{N}%
}}}}
\put(2127,-4823){\makebox(0,0)[lb]{\smash{{\SetFigFont{12}{14.4}{\rmdefault}{\mddefault}{\updefault}{H}%
}}}}
\put(2127,-5303){\makebox(0,0)[lb]{\smash{{\SetFigFont{12}{14.4}{\rmdefault}{\mddefault}{\updefault}{N}%
}}}}
\put(1659,-814){\makebox(0,0)[lb]{\smash{{\SetFigFont{12}{14.4}{\rmdefault}{\mddefault}{\updefault}{PS}%
}}}}
\put(1659,-2231){\makebox(0,0)[lb]{\smash{{\SetFigFont{12}{14.4}{\rmdefault}{\mddefault}{\updefault}{PS}%
}}}}
\put(1659,-3648){\makebox(0,0)[lb]{\smash{{\SetFigFont{12}{14.4}{\rmdefault}{\mddefault}{\updefault}{PS}%
}}}}
\put(1659,-5055){\makebox(0,0)[lb]{\smash{{\SetFigFont{12}{14.4}{\rmdefault}{\mddefault}{\updefault}{PS}%
}}}}
\put(111,-2516){\makebox(0,0)[lb]{\smash{{\SetFigFont{12}{14.4}{\rmdefault}{\mddefault}{\updefault}{start}%
}}}}
\put(1190,-403){\makebox(0,0)[lb]{\smash{{\SetFigFont{12}{14.4}{\rmdefault}{\mddefault}{\updefault}{M}%
}}}}
\put(1190,-3243){\makebox(0,0)[lb]{\smash{{\SetFigFont{12}{14.4}{\rmdefault}{\mddefault}{\updefault}{M}%
}}}}
\put(1102,-1825){\makebox(0,0)[lb]{\smash{{\SetFigFont{12}{14.4}{\rmdefault}{\mddefault}{\updefault}{GM}%
}}}}
\put(1128,-4676){\makebox(0,0)[lb]{\smash{{\SetFigFont{12}{14.4}{\rmdefault}{\mddefault}{\updefault}{GM}%
}}}}
\put(240,384){\makebox(0,0)[lb]{\smash{{\SetFigFont{12}{14.4}{\rmdefault}{\mddefault}{\updefault}{Alice}%
}}}}
\put(1651,361){\makebox(0,0)[lb]{\smash{{\SetFigFont{12}{14.4}{\rmdefault}{\mddefault}{\updefault}{Eve}%
}}}}
\put(2847,378){\makebox(0,0)[lb]{\smash{{\SetFigFont{12}{14.4}{\rmdefault}{\mddefault}{\updefault}{Bob}%
}}}}
\put(3465,390){\makebox(0,0)[lb]{\smash{{\SetFigFont{12}{14.4}{\rmdefault}{\mddefault}{\updefault}{Probability}%
}}}}
\put(3552,-94){\makebox(0,0)[lb]{\smash{{\SetFigFont{12}{14.4}{\rmdefault}{\mddefault}{\updefault}{1/8}%
}}}}
\put(3552,-566){\makebox(0,0)[lb]{\smash{{\SetFigFont{12}{14.4}{\rmdefault}{\mddefault}{\updefault}{1/4}%
}}}}
\put(3688,-1039){\makebox(0,0)[lb]{\smash{{\SetFigFont{12}{14.4}{\rmdefault}{\mddefault}{\updefault}{0}%
}}}}
\put(3552,-1511){\makebox(0,0)[lb]{\smash{{\SetFigFont{12}{14.4}{\rmdefault}{\mddefault}{\updefault}{1/8}%
}}}}
\put(3552,-1972){\makebox(0,0)[lb]{\smash{{\SetFigFont{12}{14.4}{\rmdefault}{\mddefault}{\updefault}{1/4}%
}}}}
\put(3688,-2467){\makebox(0,0)[lb]{\smash{{\SetFigFont{12}{14.4}{\rmdefault}{\mddefault}{\updefault}{0}%
}}}}
\put(3688,-2946){\makebox(0,0)[lb]{\smash{{\SetFigFont{12}{14.4}{\rmdefault}{\mddefault}{\updefault}{1}%
}}}}
\put(3688,-3406){\makebox(0,0)[lb]{\smash{{\SetFigFont{12}{14.4}{\rmdefault}{\mddefault}{\updefault}{0}%
}}}}
\put(3552,-3879){\makebox(0,0)[lb]{\smash{{\SetFigFont{12}{14.4}{\rmdefault}{\mddefault}{\updefault}{1/4}%
}}}}
\put(3688,-4339){\makebox(0,0)[lb]{\smash{{\SetFigFont{12}{14.4}{\rmdefault}{\mddefault}{\updefault}{0}%
}}}}
\put(3688,-4835){\makebox(0,0)[lb]{\smash{{\SetFigFont{12}{14.4}{\rmdefault}{\mddefault}{\updefault}{0}%
}}}}
\put(3552,-5301){\makebox(0,0)[lb]{\smash{{\SetFigFont{12}{14.4}{\rmdefault}{\mddefault}{\updefault}{1/4}%
}}}}
\put(791,-3535){\makebox(0,0)[lb]{\smash{{\SetFigFont{10}{12.0}{\rmdefault}{\mddefault}{\updefault}{1/2}%
}}}}
\put(780,-4361){\makebox(0,0)[lb]{\smash{{\SetFigFont{10}{12.0}{\rmdefault}{\mddefault}{\updefault}{1/2}%
}}}}
\put(796,-714){\makebox(0,0)[lb]{\smash{{\SetFigFont{10}{12.0}{\rmdefault}{\mddefault}{\updefault}{1/2}%
}}}}
\put(791,-1531){\makebox(0,0)[lb]{\smash{{\SetFigFont{10}{12.0}{\rmdefault}{\mddefault}{\updefault}{1/2}%
}}}}
\put(2233,-105){\makebox(0,0)[lb]{\smash{{\SetFigFont{10}{12.0}{\rmdefault}{\mddefault}{\updefault}{1}%
}}}}
\put(2471,-592){\makebox(0,0)[lb]{\smash{{\SetFigFont{10}{12.0}{\rmdefault}{\mddefault}{\updefault}{1}%
}}}}
\put(2471,-1054){\makebox(0,0)[lb]{\smash{{\SetFigFont{10}{12.0}{\rmdefault}{\mddefault}{\updefault}{1}%
}}}}
\put(2222,-1536){\makebox(0,0)[lb]{\smash{{\SetFigFont{10}{12.0}{\rmdefault}{\mddefault}{\updefault}{1}%
}}}}
\put(2466,-2013){\makebox(0,0)[lb]{\smash{{\SetFigFont{10}{12.0}{\rmdefault}{\mddefault}{\updefault}{1}%
}}}}
\put(2466,-2480){\makebox(0,0)[lb]{\smash{{\SetFigFont{10}{12.0}{\rmdefault}{\mddefault}{\updefault}{1}%
}}}}
\put(2217,-2962){\makebox(0,0)[lb]{\smash{{\SetFigFont{10}{12.0}{\rmdefault}{\mddefault}{\updefault}{1}%
}}}}
\put(2445,-3428){\makebox(0,0)[lb]{\smash{{\SetFigFont{10}{12.0}{\rmdefault}{\mddefault}{\updefault}{1}%
}}}}
\put(2450,-3900){\makebox(0,0)[lb]{\smash{{\SetFigFont{10}{12.0}{\rmdefault}{\mddefault}{\updefault}{1}%
}}}}
\put(2217,-4372){\makebox(0,0)[lb]{\smash{{\SetFigFont{10}{12.0}{\rmdefault}{\mddefault}{\updefault}{1}%
}}}}
\put(2445,-4833){\makebox(0,0)[lb]{\smash{{\SetFigFont{10}{12.0}{\rmdefault}{\mddefault}{\updefault}{1}%
}}}}
\put(2424,-5310){\makebox(0,0)[lb]{\smash{{\SetFigFont{10}{12.0}{\rmdefault}{\mddefault}{\updefault}{1}%
}}}}
\put(874,-1086){\makebox(0,0)[lb]{\smash{{\SetFigFont{12}{14.4}{\rmdefault}{\mddefault}{\updefault}{H}%
}}}}
\put(865,-3941){\makebox(0,0)[lb]{\smash{{\SetFigFont{12}{14.4}{\rmdefault}{\mddefault}{\updefault}{N}%
}}}}
\put(392,-1817){\makebox(0,0)[lb]{\smash{{\SetFigFont{10}{12.0}{\rmdefault}{\mddefault}{\updefault}{1/2}%
}}}}
\put(386,-3189){\makebox(0,0)[lb]{\smash{{\SetFigFont{10}{12.0}{\rmdefault}{\mddefault}{\updefault}{1/2}%
}}}}
\put(1778,-3497){\makebox(0,0)[lb]{\smash{{\SetFigFont{10}{12.0}{\rmdefault}{\mddefault}{\updefault}{1/4}%
}}}}
\put(1778,-3805){\makebox(0,0)[lb]{\smash{{\SetFigFont{10}{12.0}{\rmdefault}{\mddefault}{\updefault}{1/4}%
}}}}
\put(1784,-2375){\makebox(0,0)[lb]{\smash{{\SetFigFont{10}{12.0}{\rmdefault}{\mddefault}{\updefault}{1/4}%
}}}}
\put(1771,-2073){\makebox(0,0)[lb]{\smash{{\SetFigFont{10}{12.0}{\rmdefault}{\mddefault}{\updefault}{1/4}%
}}}}
\put(1778,-989){\makebox(0,0)[lb]{\smash{{\SetFigFont{10}{12.0}{\rmdefault}{\mddefault}{\updefault}{1/4}%
}}}}
\put(1771,-656){\makebox(0,0)[lb]{\smash{{\SetFigFont{10}{12.0}{\rmdefault}{\mddefault}{\updefault}{1/4}%
}}}}
\put(1791,-4896){\makebox(0,0)[lb]{\smash{{\SetFigFont{10}{12.0}{\rmdefault}{\mddefault}{\updefault}{1/4}%
}}}}
\put(1784,-5255){\makebox(0,0)[lb]{\smash{{\SetFigFont{10}{12.0}{\rmdefault}{\mddefault}{\updefault}{1/4}%
}}}}
\put(1303,-4838){\makebox(0,0)[lb]{\smash{{\SetFigFont{10}{12.0}{\rmdefault}{\mddefault}{\updefault}{1/2}%
}}}}
\put(1303,-4511){\makebox(0,0)[lb]{\smash{{\SetFigFont{10}{12.0}{\rmdefault}{\mddefault}{\updefault}{1/2}%
}}}}
\put(1284,-3420){\makebox(0,0)[lb]{\smash{{\SetFigFont{10}{12.0}{\rmdefault}{\mddefault}{\updefault}{1/2}%
}}}}
\put(1265,-3080){\makebox(0,0)[lb]{\smash{{\SetFigFont{10}{12.0}{\rmdefault}{\mddefault}{\updefault}{1/2}%
}}}}
\put(1310,-1996){\makebox(0,0)[lb]{\smash{{\SetFigFont{10}{12.0}{\rmdefault}{\mddefault}{\updefault}{1/2}%
}}}}
\put(1316,-1669){\makebox(0,0)[lb]{\smash{{\SetFigFont{10}{12.0}{\rmdefault}{\mddefault}{\updefault}{1/2}%
}}}}
\put(1258,-604){\makebox(0,0)[lb]{\smash{{\SetFigFont{10}{12.0}{\rmdefault}{\mddefault}{\updefault}{1/2}%
}}}}
\put(1233,-213){\makebox(0,0)[lb]{\smash{{\SetFigFont{10}{12.0}{\rmdefault}{\mddefault}{\updefault}{1/2}%
}}}}
\put(1650,-123){\makebox(0,0)[lb]{\smash{{\SetFigFont{12}{14.4}{\rmdefault}{\mddefault}{\updefault}{SM}%
}}}}
\put(1669,-1534){\makebox(0,0)[lb]{\smash{{\SetFigFont{12}{14.4}{\rmdefault}{\mddefault}{\updefault}{SM}%
}}}}
\put(1630,-2926){\makebox(0,0)[lb]{\smash{{\SetFigFont{12}{14.4}{\rmdefault}{\mddefault}{\updefault}{SM}%
}}}}
\put(1669,-4357){\makebox(0,0)[lb]{\smash{{\SetFigFont{12}{14.4}{\rmdefault}{\mddefault}{\updefault}{SM}%
}}}}
\end{picture}%

\end{center}
\caption{Tree of actions in the presence of eavesdropping. }
\label{ENhfig2}
\end{figure}
\end{center}

Fig.~(\ref{ENhfig2}) shows the tree of actions by Alice, Eve and Bob. The column to the right shows 
the chance that Bob get the correct message according to that path. There are five scenarios out of 
twelve possible scenarios where Bob has no chance to get the correct message, 
one scenario to get the message correctly with certainty and six different scenarios to get the correct message with 
certain probability. Probability to detect eavesdropping when hiding states will be 95.3125$\%$, higher 
than 85.9375$\%$ when Alice decide not to hide the states. The probability that Bob can't 
detect the presence of eavesdropping will be 18.75$\%$. Alice and Bob can easily 
detect the eavesdropping by communicating a sufficient number of bits.

\section{Variations to the Transmission Protocol}
\label{sec7}

Increasing the number of possible action taken by Alice and so by Bob will increase 
the confusion for an eavesdropper in preparing a superposition to be resent to Bob. This will 
increase the chance to detect the presence of eavesdropping. For example, Alice might increase her
choices by deciding to:
\begin{itemize}
\item[-] send the data directly without applying any operations. This allows Alice to send variable size
message by sending one or two bits at a time.
\item[-] encode the message as the Bell basis. 
$\frac{1}{{\sqrt 2 }}\left( {\left| {00} \right\rangle  \pm \left| {11} \right\rangle } \right),\frac{1}{{\sqrt 2 }}\left( {\left| {01} \right\rangle  \pm \left| {10} \right\rangle } \right)$. 
The 2-qubit register will represent the message instead of the oracle.
\item[-] send Bell basis as dummy data to confuse Eve.
\item[-] hide or not to hide the 
basis $\frac{1}{{\sqrt 2 }}\left( {\left| {00} \right\rangle  + \left| {11} \right\rangle } \right),\frac{1}{{\sqrt 2 }}\left( {\left| {01} \right\rangle  + \left| {10} \right\rangle } \right)$ 
to encode 0 (00 or 11) or 1(01 or 10).
\end{itemize}

\section{Conclusion}

Increasing the security of a communication channel is possible by hiding an encrypted 
message in hidden quantum states. Process of hiding and showing hidden states is fast 
since it has constant complexity. Any eavesdropper will face two problems: 
to detect the encrypted message correctly and to decrypt the message. 
By hiding the message and the eavesdropper applies direct measurement, the message will be 100$\%$ secure. If 
the eavesdropper process the superposition, the message will be 75$\%$ secure, and the 
communicating parties have a chance of 81.25$\%$ to detect the presence of eavesdropping. 
Increasing the securing of the communication is possible by applying 
simple variations to the proposed method.

\label{sec8}

%\bibliography{mybib}
%\bibliographystyle{plain}

\end{document}